\documentclass[cameraready]{Interspeech}
\newcommand\blfootnote[1]{%
  \begingroup
  \renewcommand\thefootnote{}\footnote{#1}%
  \addtocounter{footnote}{-1}%
  \endgroup
}

\title{Investigating Human-Model Discrepancies in Speech Quality Assessment \\ via Acoustic and Prosodic Perturbations}

\author[affiliation={1,*}]{Masato}{Takagi}
\author[affiliation={2}]{Masaya}{Kawamura}
\author[affiliation={2}]{Reo}{Shimizu}
\author[affiliation={2}]{Yuma}{Shirahata}
\address{
    $^1$ Nagoya Institute of Technology, Japan \\
    $^2$ LY Corporation, Japan
}

\email{m.takagi.938@stn.nitech.ac.jp, {kawamura.masaya, reshimiz, yuma.shirahata}@lycorp.co.jp}


\newcommand{\blue}[1]{\textcolor{black}{#1}}

\usepackage{comment}
\usepackage{cite}
\usepackage{here}

\newcommand{\kkedit}[1]{\textcolor{black}{#1}}
\newcommand{\sedit}[1]{\textcolor{black}{#1}}
\begin{document}

\maketitle

% # Abstract
% - SSLベースのMOS予測モデルにおいて音響的な側面以外の評価傾向は不明確
% - A~C群の摂動内容
% - 仮説H1~H3とその結果

\begin{abstract}
% 1000文字を超えていたので修正
Mean opinion score (MOS) prediction models 
%based on self-supervised learning representations
are widely used as proxy metrics in text-to-speech (TTS) research,
yet their ability to capture quality differences
beyond acoustic fidelity remains unclear.
%We investigate this via controlled perturbations on Japanese natural speech: acoustic degradation, prosodic (pitch-accent) errors, 
\sedit{We investigate this via controlled perturbations on speech: acoustic degradation, prosodic errors,}
%\blue{and speaker-characteristic variations (fundamental frequency (F0) and speaking rate) manipulation.}
\sedit{and manipulation of speaker-specific characteristics such as pitch and speaking rate.}
\sedit{We obtained MOS predictions for these speech samples from both human listeners and the model, and analyzed the differences in their perceptual characteristics.}
% Subjective ratings from native Japanese listeners and outputs of six MOS prediction models are compared.
Results show that most models track acoustic degradation well,
while all are insensitive to \sedit{prosodic} errors despite large subjective score drops.
For speaker characteristics, models exhibit a double dissociation:
strong mean fundamental frequency ($F_0$) biases absent in human ratings, yet insensitivity to
speaking rate and $F_0$ variability that humans notice.
These findings highlight limitations of scalar MOS prediction beyond \sedit{acoustic fidelity}.
%signal-level quality.
\end{abstract}

% ================================================================================
% 論文構成（コメントアウトのたたき台）
% 目的: 従来擬似MOSモデルの限界を「音響的劣化 vs. 韻律/言語的不自然さ」の観点で体系的に分析
%
% 0. Abstract
%   - 背景: 擬似MOSは運用で人間評価と乖離する事例がある
%   - 目的: 擬似MOSが何に敏感/鈍感かを実験的に分析
%   - 方法: 音響劣化/韻律・言語的誤りを制御摂動し、主観評価(自然性/韻律)と擬似MOSを比較
%   - 結果: H1–H4の要約（音響劣化は追随・韻律誤りは見落とし等）
%   - 結論: 単一スカラー評価の限界と多次元・局所性評価の必要性
%
% 1. Introduction
%   - MOS予測モデル(SSQA)の普及と運用上の乖離問題
%   - 既存研究の示唆（韻律/多様性/局所誤りでの乖離報告）
%   - 本研究の狙い: 今回やってる実験（A群、B群とか）について、何が今の擬似MOSで測れて、何が測れないのかを分析して、今後どのようなモデルが必要かを示唆
%
% 2. Related Work
%   - MOS予測モデル/ベンチマーク（VoiceMOS, MOS-Bench/SHEET等）
%   - 韻律・局所品質・フレームレベル評価の議論
%   - どういうデータセット、アーキテクチャなのかは触れた方が良さそう
%
% 3. Problem Formulation / Hypotheses
%   - 仮説H1〜H4を明示
%   - 「音響品質」と「韻律自然性」の分離評価の必要性
%
% 4. Experimental Setting
%   4.1 Data & Model
%     - 自然音声+制御摂動の方針（TTSモデル混入の排除、SiFi-GANとか）
%     - 話者・発話数・摂動種別（A:音響劣化 / B:韻律誤り / C:ピッチ/長さ操作 / D:感情）
%     - 各群のサンプル数・条件（必要なら表で整理）
%   4.2 Subjective Evaluation
%     - 5段階MOSの実施方法
%     - 評価指示（自然性MOS/韻律MOS）
%     - 被験者数・信頼区間算出
%   4.3 Objective / MOS prediction models
%     - どのようなpretrained modelを使用したか
%     - 使用モデル一覧（NISQA/UTMOS/UTMOSv2/SSLMOS/SHEET/DNSMOS/ChunkMOS等）
%     - 参照あり/なしの客観指標（PESQ/STOI/SI-SNR, $F_0$距離, SWR, WER 等）、入れなくても良いかも
%
% 5. Results
%   - A群
%   - B群
%   - C群
%
% 6. Discussion
%   どういう評価設計が必要なのか
%
% 7. Conclusion
%   - いつもの感じで
%   - 将来課題
% ================================================================================

\section{Introduction}
\label{sec:introduction}
\blfootnote{* Work done during a part-time job at LY Corporation.}
% # Introduction
% ## 背景
% - TTSモデルの性能が向上し、荒い音響劣化は減少してきている
% - 品質の差は韻律自然性・アクセント精度・話者固有特性などの細かい側面の評価に移行してきており、これらを信頼性高く評価することが重要になっている
% - 音声の品質評価ではMOS[ITU-T P.800]がスタンダードとして使われてきた
% - 一方でコスト・時間の問題から自動評価（SSQA）が求められてきた
%   - 特にSSL事前学習モデルを活用した手法(SSLMOS, UTMOS)がVoiceMOS Challengeで高い性能を示し、MOS予測の精度は大きく向上した。

% ## 問題提起 — MOS予測モデルの感度バイアス
% - 実際のTTSの研究ではMOS予測スコアを人間の評価の代わりとして用いるケースが見られる
% - しかしMOS予測モデルが人間の知覚と乖離するケースが複数の研究から示唆されている
% - 具体的には...
%   - 単一スカラーMOSの限界:音響品質・韻律・明瞭性等の多次元を1つのスコアに集約する根本的問題。多次元評価や他指標比較の試みが始まっている。[Maguer+ 2024] [Cooper+ 2024] 
%   - 韻律・アクセントへの鈍感さ: 韻律特徴量のMOS予測への寄与は限定的 [Vioni+ 2023]、SSLモデルはピッチ情報の抽出が不十分 [Shi+ 2024]、韻律多様性の評価で人間との相関が低い [Yang+ 2025]
%     - 日本語TTSのアクセント誤り検出には人間の介入が必要。[Fujii+ 2022]
%   - 話者特性バイアス: 主観評価自体が性別・年齢等の話者属性に関するバイアスを持つ。[Suda+ 2024]
%     - MOS予測モデルはTTS・VCの聴取実験データから人間の自然性を模倣するよう学習されている。
%     - そのため、話者固有の属性を品質劣化として誤認する可能性がある。
% - 一方で「MOS予測モデルがどの品質側面に敏感で品質側面に鈍感か」を、音響劣化や韻律的不自然さを制御可能な条件下で分析した研究はあまりない
%   - 本研究では、日本語(ピッチアクセント言語: アクセント誤りが語義を変えうる)に焦点を当てて、自然音声に制御された摂動を施すことで各品質側面への感度を独立に評価し、
% - MOS予測モデルの感度バイアスを明らかにする

% TODO:TTS周りの参考文献を入れる
Modern text-to-speech (TTS) systems~\cite{valle,naturalspeech3,f5tts} have reached a level of quality
that narrows the gap between synthesized and natural speech.
As coarse acoustic artifacts become less prevalent,
the quality differentiators increasingly lie in fine-grained aspects such as prosodic naturalness, accentuation accuracy, and speaker-specific characteristics~\cite{cosyvoice2,seedtts}.
Therefore evaluating these subtle differences reliably is  essential for continued progress in TTS development.

%Human listening tests based on the Mean Opinion Score (MOS) remain the gold standard for speech quality assessment~\cite{ITU_P800}.
Since speech quality is fundamentally subjective and the ultimate end-users of speech systems are humans,
listening tests based on the Mean Opinion Score (MOS) remain the gold standard for quality assessment~\cite{ITU_P800}.
However, such tests are costly, time-consuming~\cite{cooper2024review}, and difficult to reproduce.
To address these limitations, automatic MOS prediction models have been actively developed.
In particular, approaches that leverage self-supervised learning (SSL) representations,
such as SSL-MOS~\cite{sslmos} and UTMOS~\cite{utmos}, have demonstrated high prediction accuracy in the VoiceMOS Challenge series~\cite{vmc2024}.

% Yet a growing body of evidence suggests that these models do not uniformly capture all dimensions of speech quality~\cite{lemaguer2024limits,vioni2023prosody}.
% Despite these indications, MOS prediction scores are increasingly used
% as direct substitutes for human evaluation in TTS research,
% making it critical to understand their limitations.
Yet a growing body of evidence suggests that these models do not uniformly capture all dimensions of speech quality
%~\cite{lemaguer2024limits,vioni2023prosody,williams20_odyssey},
even as \blue{model-predicted MOS values} are increasingly used as direct substitutes for human evaluation \blue{including across diverse languages~\cite{f5tts,cosyvoice2}}.
% スペース確保のため(i), (ii), (iii) の形式で圧縮
Three concerns stand out:
(i)~\textbf{Information loss from scalarisation}---collapsing multidimensional quality into a single scalar inevitably discards information~\cite{lemaguer2024limits,cooper2024review,universa,ttsds2};
(ii)~\textbf{Limited sensitivity to prosody}---prosodic and linguistic cues contribute only marginally to prediction \blue{and correlate poorly with human prosody judgments~\cite{vioni2023prosody, yang2025prosodydiversity}}, 
SSL representations carry limited pitch information~\cite{shi2024pssqa},
and detecting accent errors in Japanese TTS still requires human intervention~\cite{fujii2022apsipa};
\blue{(iii)~\textbf{Speaker-dependent bias}---some speakers consistently receive high MOS while others receive low MOS regardless of the TTS system~\cite{williams20_odyssey}}\sedit{.}
%indicating that speaker-inherent characteristics influence perceived quality, yet which characteristics drive these judgments and whether prediction models appropriately reflect them remain unclear.}
% the quality achieved by certain speakers is consistent regardless of the synthesis system~\cite{williams20_odyssey}, suggesting that models trained on such ratings may inherit speaker-dependent biases.
\sedit{Collectively, these concerns raise a fundamental question: whether MOS prediction models are sensitive to the same perceptual dimensions that underlie human judgments?}
% TODO:表現を修正
% Despite these findings, it remains unclear which quality dimensions MOS prediction models capture and which they miss,
% as few studies have directly tested the sensitivity of these models along individual quality dimensions under controlled conditions.
%Despite these findings, how MOS prediction models behave relative to human perception along individual quality dimensions has not been sufficiently compared.

To address this \sedit{question},
we systematically compare human MOS ratings and model-predicted scores through controlled acoustic--prosodic perturbations applied to speech.
This design enables independent evaluation of model sensitivity along each quality dimension.
We focus on Japanese, a pitch-accent language in which accent errors can alter lexical meaning,
making prosodic accuracy particularly critical for quality assessment.
Our hypotheses and experimental design are detailed in Sections~\ref{sec:hypotheses} \sedit{and~\ref{ssec:design}, respectively}.

% \section{Related Work}
% \label{sec:related_work}
% Related Workは Experimentsの中へ含めるので省略

\section{Hypotheses}
\label{sec:hypotheses}

% # Hypotheses
% - 自然音声への制御された摂動とTTS合成時のアクセント操作を組み合わせることで、各品質次元へのモデル感度を独立に調査する
% ## H1: 音響劣化への高感度さの確認
% - 主張: MOS予測モデルは音響劣化に対して主観評価と高い相関を示す
% - 根拠: 既存モデルの学習データはノイズ・コーデック劣化・合成時のアーティファクトが豊富に含まれており音響的な品質劣化は十分に学習されていると考えられる
% - 検証: **A群（音響劣化）**: クリッピング、ノイズ付加、低ビットレートMP3圧縮
% ## H2: 韻律誤りへの鈍感さ
% - 主張: MOS予測モデルは韻律・アクセント誤りに対して鈍感である
% - 根拠: 韻律特徴のMOS予測への寄与の限定性、SSLのピッチ抽出の不十分さ、韻律の多様性評価での低相関が報告されている
% - 検証: **B群（韻律誤り）**: TTSモデルによるアクセント高低のスワップ
% - 補足: 日本語はピッチアクセント言語であり、アクセント誤りは日本語母語話者に取って知覚しやすい韻律誤りの一つであり、MOS予測モデルが言語的に意味のある韻律誤りに感度を持つかを検証するのに適している
% ## H3: 知覚的に関連する話者特性への鈍感さ (実質$F_0$と話速の変換なのでそれが伝わるように)
% - 主張: MOS予測モデルは、話者特性の変動に対して人間の知覚に影響する要因を捉えられない
% - 根拠: 学習データには話者特性の変動が品質次元として体系的にラベリングされていないため
% - 検証: **C群（ピッチ・話速操作）**:自然音声間での変化とピッチシフト・話速変換といった摂動による変化

% We evaluate the sensitivity of models along three quality dimensions
% by combining controlled perturbations of natural speech (Groups~A and~C)
% with accent manipulation during TTS synthesis (Group~B).
We formulate three hypotheses on how MOS prediction models and human listeners
differ in their sensitivity to three quality dimensions: acoustic degradation, prosodic errors, and speaker-\kkedit{characteristics}.

\noindent \textbf{H1: \sedit{Comparable} sensitivity to acoustic degradation.} 
We hypothesize that MOS prediction models \sedit{and human listeners 
will show similar sensitivity to acoustically degraded speech, 
resulting in a high correlation between their ratings.}
%exhibit high correlation with human ratings for acoustically degraded speech.
Existing models are trained on datasets that include various acoustic degradations:
the NISQA Corpus~\cite{nisqa} covers codec artifacts, noise, and packet-loss impairments,
and DNS Challenge data~\cite{dnsmos} covers additive noise conditions.
This suggests that signal-level quality loss is well represented
in their learned feature spaces.
% We test this by applying clipping, additive noise, 
% and low-bitrate MP3 compression to natural speech.

%\noindent \textbf{H2: Insensitivity to prosodic errors.}
\noindent \textbf{H2: \sedit{Reduced} sensitivity to prosodic errors.}
Based on the evidence discussed in Section~\ref{sec:introduction}~\cite{vioni2023prosody,shi2024pssqa,yang2025prosodydiversity,fujii2022apsipa},
we hypothesize that MOS prediction models \sedit{will exhibit significantly 
lower sensitivity to prosodic and accentuation errors than human listeners.}
%we hypothesize that MOS prediction models are insensitive
%to prosodic and accentuation errors.
Japanese is a pitch-accent language with a binary high--low tonal pattern
on each mora;
pitch-accent errors are among the most perceptually salient prosodic defects for native listeners,
making them well suited for testing model sensitivity
to linguistically meaningful prosodic errors.
% We test this by having a TTS model generate speech
% with deliberately flipped accent patterns.

%\noindent \textbf{H3: \sedit{Reduced} sensitivity to perceptually relevant speaker characteristics.} 
\noindent \textbf{H3: \sedit{Different} sensitivity to speaker characteristics.} 
% We hypothesize that MOS prediction models fail to capture the aspects
% of speaker characteristic variation that influence human naturalness judgments.
% As with prosodic errors (H2),
% such paralinguistic factors are not systematically labeled
% as quality dimension in existing training data,
% and models trained on a single MOS scalar are therefore unlikely to reflect them.
% We test this using Group~C, which combines unmodified natural speech from diverse speakers with artificially pitch- and speaking-rate-shifted versions of the same recordings.
\sedit{We hypothesize that MOS prediction models exhibit a different pattern of sensitivity to speaker-related characteristics compared to human listeners. 
Speaker perception involves complex interactions among pitch, speaking rate, and other voice attributes, which may not be fully captured by models trained primarily on acoustic degradation detection. 
Accordingly, model-predicted scores may not vary across speaker manipulations in the same way as human ratings.}
% Because MOS prediction models are trained primarily to detect acoustic degradation,
% they are expected to assign uniformly high scores
% to clean natural speech regardless of the speaker.
% Moreover, speaker-specific quality tends to be consistent
% regardless of the synthesis system~\cite{williams20_odyssey},
% suggesting that models may inherit such biases.
% By contrast, human listeners vary their ratings
% depending on speaker characteristics such as pitch and speaking rate.
% We test this by combining unmodified natural speech from diverse speakers
% with artificially pitch- and speaking-rate-shifted versions of the same recordings.

%The experimental procedures, including data preparation, subjective evaluation, and objective evaluation, are described in Section~\ref{sec:experiments}.

\section{Experiments}
\label{sec:experiments}

% # Experiments (実験条件)
% ## Data
% - A群
%   - クリッピング、ノイズ付加、低ビットレートMP3圧縮
% - B群
%   - TTSモデルによるアクセント高低のスワップ
% - C群
%   - C-1: 自然音声
%   - C-2: ピッチ変更
%   - C-3: 話速変更
% - 実験に使用する音声のサマリ

% ## Subjective Evaluation
% - 日本語母語話者~名、5段階MOS、3サブセットに分割

% ## Objective Evaluation
% - 各モデルのデータセットとアーキテクチャの概要

% To verify the hypotheses presented in Section~\ref{sec:hypotheses}, we conducted subjective evaluations by native Japanese speakers and objective evaluations using MOS prediction models.

% ------------------------------------------------------------------------------
\subsection{Experimental Design}
\label{ssec:design}

To verify the hypotheses presented in Section~\ref{sec:hypotheses}, we designed three groups of evaluation samples, each targeting a different quality dimension:
\begin{itemize}
    \item \textbf{Group~A (Acoustic Degradation)}: natural speech with controlled signal-level distortions (clipping, additive noise, low-bitrate MP3 compression), for testing H1.
    \item \textbf{Group~B (Prosodic Errors)}: %TTS-synthesized speech 
    \sedit{synthesized speech produced by a pitch-accent-controllable TTS system,} with deliberately flipped pitch-accent patterns, for testing H2.
    \item \textbf{Group~C (Pitch and Speaking Rate Variation)}: \sedit{(i) natural speech from speakers with diverse $F_0$ and speaking rates within the natural distribution (Group~C-1), and (ii) artificially shifted versions extending beyond the natural distribution (Groups~C-2, C-3), for testing H3.}
    %\blue{This group examines whether models track human sensitivity to $F_0$ and speaking rate by (i) comparing scores across diverse natural speakers whose characteristics vary within the natural distribution, and (ii) extending this range with artificially shifted versions that go beyond naturally occurring variation for testing H3.}
\end{itemize}
Table~\ref{tab:samples} summarizes the number of samples in each group.
We conducted subjective evaluations by native Japanese speakers and objective evaluations using MOS prediction models.
The following subsections describe the data, subjective evaluation protocol, and objective evaluation models in detail.

\begin{table}[t]
    \centering
    \caption{Number of evaluation samples per group.}
    \label{tab:samples}
    \vspace{-0.9em}
    \footnotesize
    \begin{tabular}{l l r r r r}
        \toprule
        \textbf{Group} & \textbf{Type} & \textbf{Spk.} & \textbf{Utt./Spk.} & \textbf{Cond.} & \textbf{Total} \\
        \midrule
        A        & Acoustic degrad.  & 4  & 5  & 6 & 120 \\
        B        & TTS w/ accent err.  & 4  & 10 & 3 & 120 \\
        C-1      & Natural         & 20 & 10 & 1 & 200 \\
        C-2      & Pitch shift     & 4  & 3  & 9 & 108 \\
        C-3      & Speaking rate conv.     & 4  & 3  & 9 & 108 \\
        \midrule
        \multicolumn{5}{l}{\textbf{Total}} & \textbf{656} \\
        \bottomrule
    \end{tabular}
    \vspace{-2.5em}
\end{table}
% ------------------------------------------------------------------------------
\subsection{Data}
\label{ssec:data}

\blue{For Groups~A and C,} we used the parallel100 subset of the JVS (Japanese Versatile Speech) corpus~\cite{jvs} as the source of natural speech.
This subset consists of phonetically balanced Japanese sentences read by 100 professional speakers.
\blue{\sedit{For Group~B, we} \sedit{used} synthesized speech generated by a TTS model.}

%Evaluation samples were prepared by applying controlled perturbations to the JVS recordings (Groups~A and~C) and by synthesizing speech with a proprietary TTS model (Group~B).

\subsubsection{Group~A: Acoustic Degradation}
\label{sssec:group_a}

Group~A comprises acoustically degraded versions of recordings from four speakers (two male, two female) selected from the JVS corpus, with five utterances per speaker.
% The unprocessed originals of these utterances are included in Group~C-1 (Section~\ref{sssec:group_c1}).
%\sedit{In addition to the original clean recordings,} 
Six degradation conditions were applied:
\begin{itemize}
    \item \textbf{Clipping (light)}: amplified by +20\,dB, hard-clipped, and amplitude-normalized.
    \item \textbf{Clipping (heavy)}: amplified by +40\,dB, hard-clipped, and amplitude-normalized.
    \item \textbf{Pink noise (light)}: additive pink noise at 20\,dB signal-to-noise ratio (SNR).
    \item \textbf{Pink noise (heavy)}: additive pink noise at 10\,dB SNR.
    \item \textbf{MP3 16\,kbps}: encoded to MP3 at a constant bitrate of 16\,kbps and decoded back to WAV.
    \item \textbf{MP3 8\,kbps}: encoded to MP3 at a constant bitrate of 8\,kbps and decoded back to WAV.
\end{itemize}

\noindent The four speakers used in Group~A were selected based on having the highest and lowest mean $F_0$ within each gender category, and are inherently included in the 20~speakers of Group~C-1; therefore their unprocessed originals are not counted separately.

\subsubsection{Group~B: Prosodic (Accent) Errors}
\label{sssec:group_b}

% NANSYの参考文献を入れる
% \if 0
% For Group~B, we use NANSY-TTS~\cite{nansy} as the TTS.
% To control the accent, we modified the model to take phoneme and prosodic label sequences as text input.
% This TTS model was trained on the interanl Japanese dataset with manually-annotated phonemic and
% prosodic labels, totaling 207.96 hours.
% To get the prosodic labels from grapheme, we use  DNN-based prosodic label prediction model~\cite{park2022accent}.
% % prosodic lable prediction 引用
% [1] Byeongseon Park, Ryuichi Yamamoto, and Kentaro Tachibana, “A
% unified accent estimation method based on multi-task learning for
% Japanese text-to-speech.,” in Proc. Interspeech, 2022, pp. 1931–1935.
% \fi

Group~B consists of synthesized speech produced by a NANSY-TTS model~\cite{nansy} at 24\,kHz for four speakers (two male, two female), with 10~utterances per speaker per condition.
The model was modified to accept phoneme and prosodic label sequences as text input, and was trained on an internal Japanese dataset with manually annotated phonemic and prosodic labels, totaling 207.96~hours.
During inference, prosodic labels were predicted from grapheme sequences using a deep neural network-based prosodic label prediction model~\cite{park2022accent}.
To introduce controlled prosodic errors, each sentence was segmented into accentual phrases and a subset of accentual phrases was selected with a given probability.
Within each selected accentual phrase, the binary pitch-accent labels (high $\leftrightarrow$ low) were swapped without violating the accent-type constraints, causing the model to generate locally incorrect accent patterns.
Three conditions were defined based on the accentual-phrase selection probability:
\begin{itemize}
    \item \textbf{High}: 80--90\,\% of accentual phrases selected for accent swapping.
    \item \textbf{Low}: 10--20\,\% of accentual phrases selected for accent swapping.
    \item \textbf{None}: no accent swapping.
\end{itemize}
The \textbf{None} condition serves as the baseline for the TTS model's inherent quality.
% Summary of Evaluation Samples に移動
%Note that this baseline is synthesized speech rather than natural speech, and thus differs in nature from the baselines of Groups~A and~C.

\subsubsection{Group~C: Pitch and Speaking Rate Manipulation}
\label{sssec:group_c}

Group~C consists of unmodified natural speech and natural speech with pitch or speaking rate perturbations.

% 実験結果での話の展開を踏まえて、C群の分けを変更
% C-1: 旧 C-i-2の自然音声
% C-2: 旧 C-i-1のピッチ変更音声
% C-3: 旧 C-ii-1の話速変更音声

%セクションが深すぎるので修正済み
%\paragraph{Group~C-1 (Natural Speech).}
\label{sssec:group_c1}
\noindent\textbf{Group~C-1 (Natural Speech)}
Twenty speakers were selected from the JVS corpus by a greedy algorithm that maximized the diversity of mean $F_0$ across speakers, with 10~utterances per speaker. 
%This group provides a baseline of natural inter speaker variation in pitch and speaking rate, against which the effects of artificial manipulation in Groups~C-2 and C-3 can be compared.
%The four speakers used in Groups~A, C-2, and C-3 were selected based on having the highest and lowest mean $F_0$ within each gender category, and are therefore inherently included in the 20~speakers.

%\paragraph{Group~C-2 (Pitch Shift).}
\label{sssec:group_c2}
\noindent\textbf{Group~C-2 (Pitch Shift)}
The $F_0$ of four speakers' utterances (two male, two female; three utterances each) was scaled by factors of $\{0.5, 0.7, 0.8, 0.9, 1.0, 1.1, 1.2, 1.5, 2.0\}$, followed by analysis-synthesis with SiFi-GAN~\cite{sifigan}. The $1.0\times$ condition involves analysis-synthesis without pitch modification, and thus measures the degradation introduced by the vocoder itself.
%In preliminary experiments, SiFi-GAN produced less buzzy artifacts and lower overall degradation than WORLD~\cite{world} for pitch shifting, and was therefore adopted for this group.

%\paragraph{Group~C-3 (Speaking Rate Conversion).}
\label{sssec:group_c3}
\noindent\textbf{Group~C-3 (Speaking Rate Conversion)}
The speaking rate of four speakers' utterances (two male, two female; three utterances each) was also scaled by factors of $\{0.5, 0.7, 0.8, 0.9, 1.0, 1.1, 1.2, 1.5, 2.0\}$, followed by analysis-synthesis with WORLD~\cite{world}. The $1.0\times$ condition involves analysis-synthesis without rate modification, and thus measures the degradation introduced by the vocoder itself.
%In preliminary experiments, WORLD yielded better quality than SiFi-GAN for speaking-rate conversion, and was therefore adopted for this group.

\noindent Groups~C-2 and C-3 use the same four speakers as Group A.

% The four speakers used in Groups~A, C-2, and C-3 were selected based on having the highest and lowest mean $F_0$ within each gender category, and are inherently included in the 20~speakers of Group~C-1; therefore their unprocessed originals are not counted separately.

% 消す方向で
% \subsubsection{Summary of Evaluation Samples}
% \label{sssec:samples_summary}

% Table~\ref{tab:samples} summarizes the number of samples in each group.
% The four speakers used in Groups~A, C-2, and C-3 were selected based on having the highest and lowest mean $F_0$ within each gender category, and are inherently included in the 20~speakers of Group~C-1; their unprocessed originals are therefore not counted separately.
% Note that the baseline of Group~B (the \textit{none} condition) is synthesized speech rather than natural speech, and thus differs in nature from the baselines of Groups~A and~C.
% Note that the baseline of Group~B is synthesized speech, not natural speech.
% The $1.0\times$ conditions in Groups~C-2 and C-3 (vocoder-only analysis-synthesis) are included in the condition counts.

\begin{table*}[t]
\centering
\caption{Group~A results: human MOS and MOS prediction model outputs (95\,\% CI).}
\vspace{-0.5em}
\label{tab:results_a_all}
\begin{tabular}{l c c c c c c c}
    \toprule
    \textbf{Condition} & \textbf{Human MOS} & \textbf{SHEET-MB} & \textbf{SHEET-BV} & \textbf{UTMOS} & \textbf{UTMOSv2} & \textbf{NISQA} & \textbf{DNSMOS} \\
    \midrule
    Natural   & $3.49 \pm 0.09$ & $4.63 \pm 0.01$ & $3.10 \pm 0.22$ & $3.35 \pm 0.29$ & $3.65 \pm 0.16$ & $4.56 \pm 0.19$ & $3.81 \pm 0.08$ \\
    Clipping (light)   & $1.73 \pm 0.04$ & $3.53 \pm 0.19$ & $1.87 \pm 0.10$ & $1.90 \pm 0.15$ & $2.56 \pm 0.12$ & $2.26 \pm 0.20$ & $3.40 \pm 0.11$ \\
    Clipping (heavy)   & $1.12 \pm 0.04$ & $1.73 \pm 0.15$ & $1.39 \pm 0.02$ & $1.23 \pm 0.00$ & $1.92 \pm 0.10$ & $1.26 \pm 0.04$ & $2.55 \pm 0.06$ \\
    Pink noise (light) & $1.74 \pm 0.11$ & $3.47 \pm 0.08$ & $2.58 \pm 0.20$ & $2.89 \pm 0.20$ & $2.82 \pm 0.15$ & $3.15 \pm 0.20$ & $3.13 \pm 0.07$ \\
    Pink noise (heavy) & $1.57 \pm 0.09$ & $2.92 \pm 0.15$ & $1.69 \pm 0.07$ & $1.53 \pm 0.01$ & $2.19 \pm 0.16$ & $2.24 \pm 0.17$ & $2.67 \pm 0.06$ \\
    MP3 16\,kbps       & $2.24 \pm 0.13$ & $4.29 \pm 0.09$ & $2.29 \pm 0.22$ & $2.66 \pm 0.31$ & $2.77 \pm 0.19$ & $2.59 \pm 0.28$ & $3.39 \pm 0.09$ \\
    MP3 8\,kbps        & $1.43 \pm 0.08$ & $3.76 \pm 0.19$ & $1.51 \pm 0.05$ & $1.61 \pm 0.15$ & $2.01 \pm 0.09$ & $1.39 \pm 0.07$ & $2.85 \pm 0.09$ \\
    \midrule
    utterance-level SRCC & - & 0.782 & 0.776 & 0.784 & 0.756 & 0.797 & 0.821 \\
    system-level SRCC    & - & 0.750 & 0.964 & 0.929 & 0.964 & 0.964 & 0.857 \\
    \bottomrule
\end{tabular}
\vspace{-2.0em}
\end{table*}

% ------------------------------------------------------------------------------
\subsection{Subjective Evaluation}
\label{ssec:subjective}

Fifteen native Japanese speakers participated in the subjective evaluation.
Each listener rated speech naturalness on a five-point scale, considering both acoustic quality and prosodic appropriateness as a single integrated score.
%A continuous slider ranging from 1 to 100 was used as the rating interface, and the collected scores were mapped to integer scores 1--5 using five equal-width bins of 20 points each for analysis.
\blue{Each listener evaluated all 656~samples, with all conditions intermixed to prevent habituation to specific degradation patterns.}
We computed MOS by averaging across listeners for each sample.
All samples were presented at 24\,kHz.
% We report 95\,\% confidence intervals for all MOS values.

% ------------------------------------------------------------------------------
\subsection{Objective Evaluation}
\label{ssec:objective}

% \subsubsection{MOS prediction models}
% \label{sssec:pseudo_mos}

All MOS prediction models were run through VERSA~\cite{versa, versatoolkit}, 
a unified evaluation toolkit that provides standardized inference scripts and pretrained
weights for reproducible assessment.
We evaluated six models.
Input audio was resampled to 16 kHz \sedit{to match the expected input format of the pretrained models.} %at 16-bit resolution before inference.

\begin{itemize}
    \item \textbf{SHEET-MB}~\cite{sslmos, sheet}: SSL-MOS architecture, which combines a self-supervised speech encoder with a linear prediction head. WavLM~\cite{wavlm} Large is used as the encoder. 
    Trained on MOS-Bench~\cite{mosbench}, a combination of eight datasets (including BVCC~\cite{bvcc}, SingMOS~\cite{singmos}, NISQA)
    covering TTS, voice conversion (VC), noisy, and telephone speech.
    \item \textbf{SHEET-BV}~\cite{sslmos, sheet}: SSL-MOS architecture with WavLM Large (same as SHEET-MB). Trained on BVCC alone (English TTS/VC).
    \item \textbf{UTMOS}~\cite{utmos}: wav2vec 2.0~\cite{wav2vec2} with data augmentation, listener-dependent modeling, and domain adaptation. 
    Trained on BVCC (English TTS/VC).
    \item \textbf{UTMOSv2}~\cite{utmosv2}: wav2vec 2.0 fused with EfficientNetV2~\cite{efficientnetv2} spectrogram features. Trained on BVCC (English TTS/VC).
    \item \textbf{NISQA}~\cite{nisqa}: Convolutional neural network (CNN) on mel-spectrograms. 
    Trained on NISQA Corpus (codec, noise, and packet-loss degradations; no TTS/VC data; German/English).
    \item \textbf{DNSMOS}~\cite{dnsmos}: CNN on log-mel spectrograms.
    Trained on DNS Challenge data (noisy/enhanced speech; no TTS/VC data; English).
\end{itemize}

\noindent
%None of the models were trained on Japanese data.

% \subsubsection{Reference-Based Metrics}
% \label{sssec:ref_metrics}
% 今回はなし

% For Groups~A, C-1-i, and C-2, where the original unprocessed recordings are available as references, we additionally computed PESQ~\cite{pesq}, STOI~\cite{stoi}, and SI-SNR~\cite{sisnr}.

\subsection{Results}
\label{sec:results}

% # Experiments (実験結果)
% ## Results
% - 主観評価: MOS（95%信頼区間）
% - 客観評価: 各MOS予測モデルの出力と（A群は）人手MOSとのSRCC（system/utterance）

% ## Group A: Acoustic Degradation（H1）
% - 目的: 音響劣化に対するモデル感度（人手MOSとの相関）を確認
% - 主観MOS: Natural 3.49に対して、劣化条件は1.12〜2.24まで有意に低下し、重症度に対して単調に低下
% - 相関（SRCC）: 多くのモデルでsystem-levelは高い一方、SHEET-MBのみ低下（0.75）
%   - 例: MP3 8 kbpsで人手1.43に対し予測3.76となり順位が崩れる
% - 学習データ vs アーキテクチャ: 同一WavLM Largeでも学習データ差でsystem-level SRCCが0.75→0.964に改善（MOS-Bench→BVCC）
%   - 一方、同一データでもSSLエンコーダ差（WavLM vs wav2vec2）ではパターンは類似（SHEET-BV 0.964 / UTMOS 0.929）
% - utterance-level: 2種のSSL-MOSで近いSRCC（0.782 vs 0.776）で、system-levelの崩れと対照的
% - 結論: H1はモデル依存で支持（概ね高相関だが例外あり）

% ## Group B: Prosodic（Accent）Errors（H2）
% - 目的: アクセント誤り（韻律）へのモデル感度を検証
% - 主観MOS: None 4.00 → Low 3.19 → High 2.16（合計-1.84）で日本語母語話者が明確
% - 予測モデル: いずれも操作にほぼ無反応（3条件間の最大差が各モデルで<0.10）
%   - 例: SHEET-MB 4.63→4.63→4.63、DNSMOS 3.82→3.81→3.83
% - 結論: H2を強く支持（SSL/非SSLとも一様に鈍感）

% ## Group C: Speaker Characteristics（H3）
% - 目的: 話者特性（平均$F_0$、$F_0$変動、発話時間/話速）が主観・客観スコアに与える影響を比較
% - 指標: 話者特性とスコアのPearson相関r（C-1〜C-3）
% - C-1（自然音声・話者差）
%   - 人手MOS: 平均$F_0$はほぼ無相関（r=-0.02）
%   - 人手MOS: 発話時間（話速）はr=-0.52で、観察上は逆U字傾向、$F_0$変動は正相関（r=0.48）
%   - モデル: 平均$F_0$に強い負のバイアス（例: DNSMOS -0.84、UTMOSv2 -0.76、SHEET-BV -0.67）で、人手にはない傾向
%   - モデル: $F_0$変動・発話時間は概ね0付近（例外: SHEET-MBの発話時間 r=0.42）
% - C-2（ピッチシフト）
%   - 人手MOS: 元$F_0$（1.0x）でピーク、上下に対称に低下（平均$F_0$との相関r=0.04）
%   - モデル: C-1と同様の平均$F_0$バイアスが残存
% - C-3（話速変換）
%   - 一部モデルで発話時間との相関が強まる（例: NISQA 0.23→0.59）
%     - メモ: SSLを使っていないCNNベースのモデルで、C-1→C-3での相関上昇が顕著
%     - 時間伸縮によって生じるセグメントあたりのスペクトル変動の減少を高品質と捉えた可能性がある
% - 結論: H3を支持（人手が影響を受ける話者特性を捉えられず、人間の評価に表れない平均$F_0$バイアスが顕著）

We report results for Groups~A--C. Subjective results are reported as MOS with 95\,\% confidence intervals. Objective results are reported for each MOS prediction model.

% ------------------------------------------------------------------------------
\subsubsection{Group~A: Acoustic Degradation (H1)}
\label{ssec:results_group_a}
% MOS-Benchで音響劣化に関する汎化性が下がっている

Table~\ref{tab:results_a_all} shows human and model-predicted MOS for each acoustic degradation condition,
along with system-level and utterance-level Spearman's rank correlation coefficients (SRCCs) between each model and human.
All degradation conditions produced human MOS values significantly lower than the natural condition (3.49), ranging from 1.12 to 2.24, with monotonic decreases as severity increased.
Most models achieved high system-level SRCC with subjective ratings.
The exception was SHEET-MB at 0.750, due to disordering in the MP3 conditions
(MP3 8\,kbps: predicted 3.76 vs.\ human 1.43).
This is consistent with the reduced generalization to acoustic degradation observed for multi-domain models in MOS-Bench~\cite{mosbench}.

Comparing the two SSL-MOS variants, SHEET-MB and SHEET-BV, which share the same architecture but differ in training data, the system-level SRCC improved from 0.750 (SHEET-MB) to 0.964 (SHEET-BV).
Meanwhile, SHEET-BV (WavLM) and UTMOS (wav2vec~2.0),
trained on the same data but with different SSL encoders,
showed similar patterns (0.964 vs.\ 0.929).
This suggests that the composition of training-data, rather than SSL architecture,
is the dominant factor governing sensitivity to acoustic degradation.
At the utterance level, however, the two SSL-MOS variants achieved comparable SRCCs (0.782 vs.\ 0.776),
indicating that multi-domain training disrupts system-level ranking
while preserving utterance-level ordering.
From these results, H1 is largely supported, with SHEET-MB as the sole exception.

% \blue{
% These results indicate that training corpora for acoustic degradation contain
% explicit quality annotations over degraded speech,
% enabling models to detect the manipulation.
% The sensitivity observed in Group~A is thus consistent with the presence of labeled supervision
% for this quality dimension.
% }

\begin{table*}[t]
\centering
\footnotesize

% ---------- A ----------
% \begin{minipage}{\textwidth}
% \end{minipage}
% \vspace{0.8em}

% ---------- B ----------
\begin{minipage}{\textwidth}
\centering
\caption{Group~B results: human MOS and MOS prediction model outputs (95\,\% CI).}
\label{tab:results_b_all}
\vspace{-1.0em}
\begin{tabular}{l c c c c c c c}
    \toprule
    \textbf{Condition} & \textbf{Human MOS} & \textbf{SHEET-MB} & \textbf{SHEET-BV} & \textbf{UTMOS} & \textbf{UTMOSv2} & \textbf{NISQA} & \textbf{DNSMOS} \\
    \midrule
    None (baseline)        & $4.00 \pm 0.07$ & $4.63 \pm 0.01$ & $3.05 \pm 0.07$ & $3.44 \pm 0.12$ & $3.56 \pm 0.08$ & $3.81 \pm 0.16$ & $3.82 \pm 0.05$ \\
    Low (swap 10-20\,\%)  & $3.19 \pm 0.09$ & $4.63 \pm 0.01$ & $3.07 \pm 0.07$ & $3.43 \pm 0.11$ & $3.62 \pm 0.06$ & $3.74 \pm 0.15$ & $3.81 \pm 0.06$ \\
    High (swap 80-90\,\%) & $2.16 \pm 0.09$ & $4.63 \pm 0.01$ & $3.09 \pm 0.07$ & $3.47 \pm 0.11$ & $3.61 \pm 0.08$ & $3.84 \pm 0.16$ & $3.83 \pm 0.05$ \\
    \bottomrule
\end{tabular}
\end{minipage}

\vspace{0.4em}

% ---------- C ----------
\begin{minipage}{\textwidth}
\centering
\caption{Pearson correlation coefficients ($r$) between speaker characteristics and scores in Group~C.}
\label{tab:results_c}
\vspace{-0.8em}
\begin{tabular}{l|rrrrrrr}
\toprule
 & \textbf{Human MOS} & \textbf{SHEET-MB} & \textbf{SHEET-BV} & \textbf{UTMOS} & \textbf{UTMOSv2} & \textbf{NISQA} & \textbf{DNSMOS} \\
\midrule

\blue{C-1 Mean $\log F_0$}     & -0.059 &  0.549 & -0.618 & -0.530 & -0.722 & -0.531 & -0.788 \\
C-1 Std. $\log F_0$      &  0.477 &  0.014 & -0.196 & -0.106 & -0.007 & -0.158 & -0.105 \\
C-1 Dur.                & -0.520 &  0.420 &  0.140 &  0.210 & -0.300 &  0.230 & -0.010 \\
\midrule
\blue{C-2 Mean $\log F_0$}     &  0.113 & -0.058 & -0.458 & -0.374 & -0.762 & -0.520 & -0.724 \\  
\midrule
C-3 Dur.                & -0.382 &  0.379 &  0.191 &  0.120 & -0.353 &  0.585 &  0.294 \\
\bottomrule
\end{tabular}
\end{minipage}

\vspace{-1.5em}
\end{table*}

% ------------------------------------------------------------------------------
\subsubsection{Group~B: Prosodic (Accent) Errors (H2)}
\label{ssec:results_group_b}

Table~\ref{tab:results_b_all} shows human and model-predicted MOS for \blue{each accent swapping condition.
Increasing the proportion of swapped accentual phrases} led to a substantial decline in human MOS:
None $4.00 \to$ Low $3.19 \to$ High $2.16$,
a total drop of 1.84 points.
Native Japanese listeners clearly perceived accent errors
and reflected them in their naturalness ratings.
In contrast, all MOS prediction models \sedit{exhibited negligible variation} %were virtually unresponsive
to the accent manipulation.
\blue{
Although human MOS dropped by 1.84 points, no model exhibited more than 0.1 points of change across the three conditions.
}
This insensitivity was uniform across both SSL-based and non-SSL models.

\blue{
Comparing Groups~A and~B highlights an asymmetry in the role of training data.
In Group~A, changing training data from MOS-Bench to BVCC
restored sensitivity to acoustic degradation,
yet in Group~B, SHEET-BV was equally insensitive to prosodic errors.
}
\sedit{This pattern suggests that altering data composition alone is insufficient to induce sensitivity to prosodic appropriateness. 
Rather, the lack of explicit supervision targeting prosodic quality may be a limiting factor, as current MOS datasets do not provide annotations that isolate prosodic appropriateness from overall acoustic quality.}
%This asymmetry indicates that for quality dimensions lacking explicit supervision,
%the problem lies not in data composition but in absence of supervisory signal itself;
%no existing dataset provides annotations that isolate prosodic appropriateness from acoustic quality.
The insensitivity to prosodic manipulation is also consistent with prior findings that SSL models have limited capacity for $F_0$ extraction~\cite{shi2024pssqa} and that prosodic features contribute only marginally to MOS predictions~\cite{vioni2023prosody}.
Therefore, H2 is strongly supported.

Additionally, all models were trained without Japanese \emph{speech} data
(SHEET-MB includes Japanese and Chinese singing data via SingMOS~\cite{singmos}, but not Japanese speech),
so language mismatch may partly explain the insensitivity observed in Group~B.
However, these models are routinely used as language-independent evaluation metrics for speech synthesis across diverse languages\blue{~\cite{f5tts, cosyvoice2, park2024},}
and the present results \sedit{highlight potential limitations of this practice.} %call that practice into question.
%F5-TTSが中国語でUTMOSを使用しているため、引用をつける
% introにも追加
%他の論文
% Analytic Study of Text-Free Speech Synthesis for Raw Audio using a Self-Supervised Learning Model (https://arxiv.org/abs/2412.03074) 日本語にUTMOS
% Cosyvoice2 は中国語音声の評価にDNSMOSを使用

% ------------------------------------------------------------------------------
\subsubsection{Group~C: Speaker Characteristics (H3)}
\label{ssec:results_group_c}

\begin{figure}
    \centering
    \vspace{-0.5em}
    \includegraphics[width=1\linewidth]{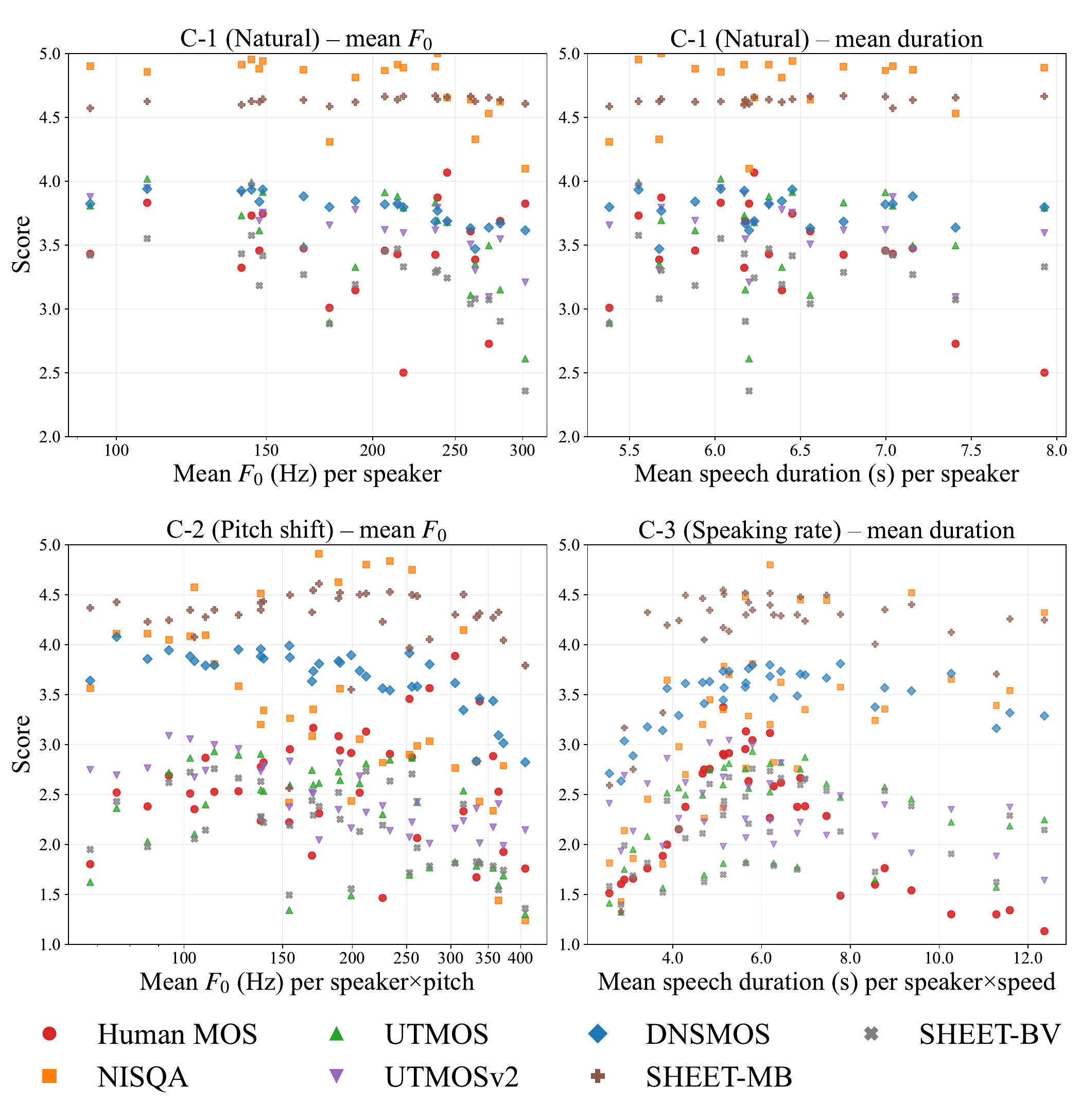}
    \vspace{-2.0em}
    \caption{Scatter plots of MOS against speaker characteristics (mean $\log F_0$, speaking rate) for Groups C-1--C-3.}
    \label{fig:2x2}
    \vspace{-2.0em}
\end{figure}

\blue{
Table~\ref{tab:results_c} shows Pearson correlation coefficients $r$ between speaker characteristics (mean $\log F_0$, standard deviation of $\log F_0$, speaking rate) and scores across Groups C-1--C-3.
Correlations are computed for both human and model-predicted MOS.
}
%分析としてmean F0, std log F0, durationを見るという文章を入れる。
%Figの説明を入れる
In Group~C-1, mean $\log F_0$ showed \sedit{a negligible} correlation with human MOS ($r = -0.059$).
\blue{We further examined correlations with mean speaking rate and variation of $\log F_0$.}
%Examining speakers whose scores deviated from the group mean $\log F_0$,
%追加でstd log F0, durationを見たところ
We found that mean speaking rate exhibited a negative correlation
with human MOS ($r = -0.520$, Fig.~\ref{fig:2x2});
since all speakers read the same sentences,
the differences reflect speaker-specific speaking rates.
The variation of $\log F_0$ (standard deviation of $\log F_0$)
was also correlated positively with human MOS ($r = 0.477$).
The models showed two contrasting patterns relative to human listeners.
For mean $\log F_0$, where humans did not show correlation,
most of the models showed strong negative correlations,
assigning higher scores to \blue{lower-$F_0$} speakers.
In contrast, for $\log F_0$ variability and speaking rate, where humans showed moderate correlations,
all models showed near zero ($\log F_0$ variability: $r=-0.20$ to $0.01$).
%These results indicate that the models were not merely insensitive to $\log F_0$ variability and speaking rate,
%but also showed biases toward mean $F_0$.
% in \blue{their predicted MOS}.
%This suggests that models may encode implicit biases from the speaker characteristics of the training data rather than capturing perceptually relevant variation.
These results suggest that the models are influenced by speaker-dependent acoustic characteristics that are not strongly associated with human perceptual ratings. This pattern is consistent with the possibility that the models internalize distributional properties of the training data rather than perceptually salient variation.

%except SHEET-MB for speaking rate ($r=0.420$).
\blue{SHEET-MB was a consistent outlier, showing a positive $F_0$ correlation ($r=0.549$) and a moderate speaking rate correlation ($r=0.420$).
This outlier behavior is plausibly explained by the SingMOS singing voice data in its training set.
In singing-voice data, high-$F_0$ voices and long duration notes are prevalent and associated with high ratings, \sedit{which may have influenced the model’s learned associations in a direction opposite to those trained solely on speech data.}
%introducing a bias in the opposite direction relative to models trained solely on speech data.
}

For Group~C-2, the overall scoring trend did not change;
% 人間が(軽い正の傾向)でモデルは変わらず負の傾向を示していてスコアの傾向は変わらないという書き方に
human scores exhibited a very weak association with $F_0$ ($r=0.113$), 
%despite perceptual sensitivity to vocoder artifacts. In contrast, 
and the models maintained the same negative association as in Group~C-1. 

%Humans showed a weakly positive correlation with $F_0$ scaling ($r=0.113$, Fig.~\ref{fig:2x2}$), despite perceptual sensitivity to vocoder artifacts. In contrast, the model continued to show a negative trend consistent with Group~C-1.

%human scores peaked at the original $F_0$ (1.0$\times$) and declined symmetrically ($r=0.113$, Fig.~\ref{fig:2x2}),
%human scores peaked at 6.0s and declined symmetrically (Fig.~\ref{fig:2x2}),
%consistent with perceptual sensitivity to vocoder artifacts. The model correlation patterns were not changed from Group~C-1.

In Group~C-3,
human scores peaked at 6.0s and declined symmetrically (Fig.~\ref{fig:2x2}), while models did not exhibit a comparably strong symmetric pattern.
Mel-spectrogram-based models showed stronger speaking rate correlations \blue{(NISQA: $r= -0.230$ to $0.585$, DNSMOS: $r= -0.010$ to $0.294$)} than C-1.
Time stretching reduces within-segment spectral variation, which these CNNs likely interpret as higher quality.

% In all cases, the bias reflects training data properties rather than
% any relationship observed in human perception.
% Models fail to capture speaker characteristics that influence human judgments, while exhibiting strong biases toward mean $\log F_0$ that are absent in human ratings.
% From these results, H3 is supported.
% Since training data labels define the boundary of what models can and cannot evaluate,
% extending supervision to prosodic appropriateness and speaker characteristics as independent quality dimensions
% may be necessary to move beyond signal-level assessment.

\sedit{Across conditions, model predictions exhibited systematic divergence from human perceptual patterns. 
While human MOS showed moderate associations with speaking rate and $\log F_0$ variability, most models instead displayed strong correlations with mean $\log F_0$, a trend largely absent in human ratings. 
These results indicate that MOS prediction models do not replicate the perceptual structure underlying human judgments of speaker characteristics, thereby supporting H3. These findings suggest that current MOS training paradigms may not adequately encode prosodic and speaker-related dimensions.}

\section{Conclusion}
\label{sec:conclusion}
% # Conclusion
% - 結果の要約
%   - 現行MOS予測モデルは音響劣化は高信頼に検出
%   - 一方で韻律（アクセント）誤りには一様に鈍感
%   - さらに、人間判断に影響する話者特性を捉えられず、平均$F_0$への強いバイアス（人間には無い）を示す
% - 単一スカラーMOS予測は信号レベル品質の指標としては有効だが、自然性の知覚的に重要な次元（韻律・話者特性など）を反映できない

% \blue{Under our evaluation setting}, MOS prediction models
% reliably tracked acoustic degradations but are uniformly insensitive
% to prosodic errors despite human MOS drops.
% For speaker characteristics, models exhibited strong mean $\log F_0$ biases absent in human MOS yet missed $\log F_0$ variability and speaking rate that listeners perceived.
% These results suggest that single \sedit{scalar} MOS prediction capture signal-level quality but not perceptually relevant dimensions.
Under controlled acoustic-prosodic perturbations, MOS prediction models reliably tracked signal-level acoustic degradations but were insensitive to linguistically meaningful prosodic errors and exhibited systematically misaligned sensitivity to speaker-related characteristics.
Collectively, these findings demonstrate that current MOS prediction models do not replicate the perceptual structure underlying human quality judgments. While human listeners integrate multiple acoustic and prosodic cues, existing models emphasize a different subset of features, leading to systematic divergence from human evaluation patterns.
%\blue{Future work should extend supervision to prosodic and speaker dimensions.}
\sedit{Future work should develop evaluation frameworks that enable MOS prediction models to capture prosodic and speaker-related quality dimensions beyond signal-level acoustic degradation.}

\if 0
Under controlled acoustic–prosodic perturbations, MOS prediction models reliably tracked signal-level acoustic degradations but were insensitive to linguistically meaningful prosodic errors and exhibited systematically misaligned sensitivity to speaker-related characteristics.
Collectively, these findings demonstrate that current MOS prediction models do not replicate the perceptual structure underlying human quality judgments. While human listeners integrate multiple acoustic and prosodic cues, existing models emphasize a different subset of features, leading to systematic divergence from human evaluation patterns.
% ---
Improving MOS prediction will therefore require closer alignment between model sensitivity and the perceptual dimensions that govern human judgments.
\fi

\newpage

\section{Generative AI Use Disclosure}
Generative AI tools were used to assist in English language editing of the manuscript.
All authors have reviewed the final version and take full responsibility for the scientific content, experimental design, results, and conclusions.

\bibliographystyle{IEEEtran}
\bibliography{mybib}

\end{document}